\documentclass{IET-Conf-Paper}

\usepackage[colorlinks=true, allcolors=blue]{hyperref}

\usepackage{eurosym}
\usepackage{subcaption}
\usepackage{tikz}
\usepackage{pgfplots}
\usepgfplotslibrary{groupplots}
\usetikzlibrary{svg.path}
\usetikzlibrary{shapes,arrows}
\usepackage{pgfplots}
\usepgfplotslibrary{fillbetween}
\usetikzlibrary{pgfplots.statistics}
\usepackage{breqn}
\usepackage[boxruled]{algorithm2e}
\usepackage{placeins}
\begin{document}

\title{Optimizing Offshore Wind Integration through Multi-Terminal DC Grids: A Market-Based OPF Framework for the North Sea Interconnectors}

\author{Bernardo Castro Valerio\ad{1,2}\corr,  Vinícius Albernaz Lacerda\ad{1}
, Marc Cheah-Mañe\ad{1} ,Pieter Gebraad\ad{2} and Oriol Gomis-Bellmunt \ad{1}}

\address{\add{1}{Centre d’Innovació Tecnològica en Convertidors
Estatics i Accionaments (CITCEA-UPC),
Barcelona, Spain
\add{2}{Youwind Renewables, Barcelona, Spain}}
\email{bernardo.castro@upc.edu}}


\begin{abstract}

Interconnecting price zones and remote renewable energy sources has emerged as a key solution to achieving climate goals. The objective of this work is to present a formulation that extends the base optimal power flow model with price zones constraints to forecast the operations of upcoming offshore wind developments integrated into a multi-terminal DC grid. A case study based on the 2030 development of the North Sea is used to exemplify the utilization of the formulation. Here, three cases are presented, one with the price as a parameter and the other two with the price as a variable dependent on power flows between price zones. The paper demonstrates that, for large power flows, it is necessary to include additional constraints beyond line limitations to accurately capture the effects of price zone exchanges.

\end{abstract}

\maketitle

\section{Introduction}

With the installation of the largest green energy power plant and the interconnection of cross-border electricity, Europe aims to achieve its climate and energy goals \cite{a2023_expert,european_commission_2017}. The North Sea cooperation has proposed a multi-purpose multi-terminal HVDC (MTDC) grid to interconnect 260 GW of offshore wind capacity by 2050 \cite{a2024_expert}. Thus, understanding how power flows between the various energy hubs and countries has become essential.

Wind integration is often based on the assumption that power injections do not interfere with the larger-scale wholesale market. However, this assumption breaks down when considering hub-level power injections, such as those proposed by the North Sea Cooperation. A fundamental distinction between a price-maker and a price-taker model must be considered, which lies in whether the influence of offers on cleared prices is accounted for \cite{makersetter}.

Reference \cite{TOSATTO2022112907} presents the first large-scale impact analysis of offshore hubs on the entire European power system and electricity market. The study employs a lossless linear model of the power system and uses the Power Transfer Distribution Factor matrix to model market behaviour. Reference \cite{MKTrnsmission} describes each market as a singular node, assuming linearity for the supply and demand curves near the equilibrium point. This assumption allows for an approximation of an incremental social cost. Both references oversimplify the electric grid:  \cite{TOSATTO2022112907}  ignores losses, while \cite{MKTrnsmission} does not consider local AC networks or converters.

A balance must be defined between model complexity and computational efficiency. This paper aims to build upon non-linear AC/DC optimal power flow (OPF) and expands it to include limitations on power flows between regions. These regions, referred to as price zones, can be modelled as singular nodes or, more importantly, as collections of AC and DC nodes, as presented in the formulation.

The main contribution of this paper is an extension of the non-linear AC/DC OPF formulation to account for power interchanges between price zones and their relationship to electricity prices.

The methodology employed to achieve this is presented in Section \ref{sec:Methodology}, providing a brief recap of the day-ahead market, defining the cost of generation, and its integration into the OPF formulation. In Section \ref{sec:Case Study}, a North Sea case study is used to exemplify the formulation.

\section{Methodology} \label{sec:Methodology}

\subsection{Day-Ahead Market}
Day-ahead and intraday are considered the backbone of the European spot market. The day-ahead market operates through blind auctions, where bid amounts and bidder identities are kept hidden from all participants. The auctions are held once a day, during which all hours of the following day are traded. Market participants provide their willingness to buy or sell a specific volume as well as block orders linking several delivery periods together \cite{EPEXSpotBasics}.

\begin{figure}[h]
    \centering
    \includegraphics[width=\columnwidth]{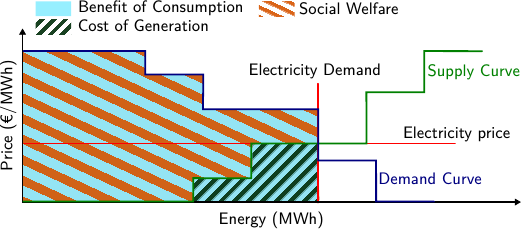}
    \caption{Generic supply and demand curves in price zones}
    \label{fig:MeriOrder}
\end{figure}

Electricity prices are based on marginal cost, which represents the cost of producing an additional megawatt-hour (MWh). Generators within a price zone are ranked by increasing marginal cost, starting with the least expensive. The market-clearing price is determined by the marginal cost of the last generator activated to meet demand, as shown in \autoref{fig:MeriOrder}.

Aggregated demand and supply curves are formed from buy and sell orders, respectively. The price is set hourly for the following day, with all producers receiving the same price corresponding to the marginal cost of the last activated plant \cite{pricesset}.

With increased interconnection electricity flows freely across borders, and more efficiently through different price zones having an effect on social welfare. The objective of price zone modelling for OPF is to establish a relationship between imports and exports and the corresponding price zone prices. For example, excessive export will result on additional power plants activating and increasing the price zone price. 

\subsection{Cost of generation}

When a price zone is modelled as a singular node, imports and exports are compressed to total net power injections defined as: 

\begin{equation} \label{eq:netPowerMarket}
    P_N=P_{G}-P_{D}  \qquad \forall  m \in \mathcal{M}
\end{equation}

Where $P_{G}$ is the total generation and $P_{D}$ the total demand. $P_N$ is positive if the price zone is exporter whereas it is is negative when is importer \cite{MKTrnsmission,peakloadThesis}. Social Welfare (SW) is  defined for each hour as the Benefit of consumption (BC) minus the cost of generation (CG) , shown in \autoref{fig:MeriOrder}, as:

\begin{equation}\label{eq:SW}
    \centering
      SW= BC_0 - CG_0
  \end{equation}
  
\noindent where $CG_0$ and $BC_0$ are obtained by integrating the area beneath the aggregated supply and demand bidding curves from zero to where supply equals demand \cite{MKTrnsmission,folk,peakloadpricing}. If the loads $P_D$ are treated as fixed parameters, $BC_0$ remains constant, and $P_D$ acts as a horizontal translation of $CG$. Therefore, to maximize SW the objective becomes to minimize CG.

\begin{equation}\label{eq:CG_obj}
    \min \sum^{\mathcal{M}}_m CG(P_N)_m
\end{equation}

\noindent where $CG_0$ is approximated using a quadratic function \eqref{eq:CG}. However, in optimisation problems,  the fixed cost can be omitted since it does not affect the optimal solution \cite{MKTrnsmission}. Thus, CG is defined as shown in \eqref{eq:CG_inc}.

\begin{subequations}
\begin{align}
     CG_0(P_N)&=\alpha P_N^2+ \beta P_N+ \text{fixed cost} \label{eq:CG} \\
     CG(P_N)&=\alpha P_N^2+ \beta P_N \label{eq:CG_inc}
\end{align}
    
\end{subequations}

The price is then determined by the marginal cost of power generation as:

\begin{subequations}
\begin{align}
    \rho &= \frac{\delta CG}{\delta P_N}= 2\alpha P_N+\beta \label{eq:price_var}\\
     \rho_{eq} &=\frac{\delta CG}{\delta P_N}(0)=\beta
\end{align}
    
\end{subequations}

where the equilibrium price $\rho_{eq}$ is defined as the price at which the aggregated supply and demand curves intersect. When $ P_N = 0$, no net power flows into or out of the price zone, and the price  $\rho $ reflects this supply-demand balance. For an exporter price zone ($ P_N > 0$ ),  $\rho$  will rise relative to the equilibrium price due to excess need for generation. Conversely, for an importer price zone ( $P_N < 0$ ),  $\rho$  will fall relative to the equilibrium price due to decreased need for generation.

To extend this strategy from single-node modelling to constrain multiple nodes within a single price zone, the generation and loads of all nodes is aggregated such that:

\begin{equation}\label{eq:PG}
    P_G = \sum^{i \in M }_i \left( P_{g_i} + \sum\gamma P_{rg_i} \right)
\end{equation}

\begin{equation}\label{eq:PD}
    P_D = \sum^{i \in M }_i L_i
\end{equation}

 \noindent where $P_{g_i}$ represents the available power by a renewable sources so that $\gamma_{rg}P_{rg}$ represents the curtailed renewable source power. Curtailment is integrated through a variable $\gamma$, where curtailment in per unit is defined as $1 - \gamma$. $P_{g_i}$ represents the active power generated by non renewable sources in the node and variable $L_i$ represents a load's consumption.

\subsection{Optimal power Flow}

The optimal power flow of AC/DC grids considers the curtailment of renewable energy sources and the power set points of converters and generators. The optimisation considers loading inputs and renewable energy source availability. It adheres to power flow relationships in AC networks, DC networks, and their interconnections while ensuring compliance with the requirements for the voltage magnitude and angle of AC nodes, the voltage magnitude of DC nodes, and the power ratings of AC branches, DC lines, generators, and converters.  A more detailed description of OPF variables and constraints can be found in \cite{Bernardo_Pyflow}. 

\begin{figure}
    \centering
    \includegraphics[width=0.9\columnwidth]{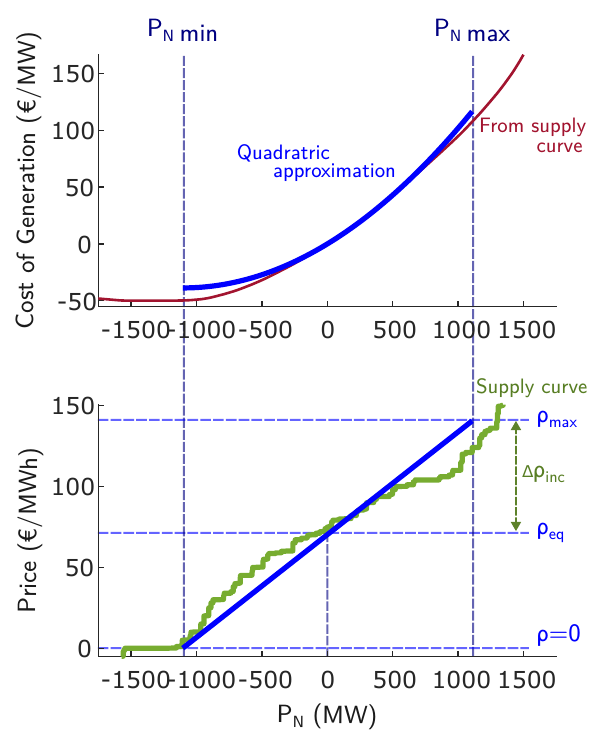}
    \caption{CG of the Netherlands for 16/09/2024 13:00-14:00 based on data from \cite{epexspot}}
    \label{fig:NL}
\end{figure}

Equations \eqref{eq:netPowerMarket},\eqref{eq:CG_inc},\eqref{eq:PG} and \eqref{eq:PD} are added as equality constraints for each price zone. Additionally, an inequality constraint for each price zone is added such that:

\begin{equation} \label{eq: PNconstraint}
    P_{N_{min}} \leq P_{N} \leq P_{N_{max}}  \quad \forall m \in \mathcal{M}
\end{equation}

\noindent where $P_{N_{min}}$ reflects the maximum import of the price zone and is defined by \eqref{eq:MPmin} where parameter $\beta$ specifies whether it corresponds to the minimum of the curve or the equilibrium point. On the other hand, the maximum export $P_{N_{max}}$ is defined by the relative to a maximum price increase ($\Delta \rho_{inc}$) to be observed. The value of $\Delta \rho_{inc}$ is selected arbitrarily within the supply curve to constrain the data range, thereby limiting the extent to which a price zone is willing to increase its price, facilitating the quadratic approximation. A visual representation of such variables is seen in \autoref{fig:NL}.

\begin{equation} \label{eq:MPmin}
   P_{N_{min}}=\begin{cases} \frac{-\beta}{2\alpha} & \beta > 0 \\
 0 & \beta \leq 0          
    \end{cases}
\end{equation}

The optimisation objective \eqref{eq:CG_obj} does not differentiate between prioritizing renewable energy sources from generators in the OPF. As a result, the outcome tends to be very similar to loss minimization objective, where generation is prioritized by proximity to the loads to reduce power loss. By not considering each price zone as a singular node but as multiple nodes, losses within the price zones are inherently involved. The higher the losses, the higher the difference between generation and load ($P_N$) of each price zone, which increases the CG. Furthermore, CG adds an element where the formulation tends to reach an equilibrium of electricity prices throughout all the price zones.

In order to prioritise offshore renewable generation, offshore nodes need to be described by external price zones. The supply curve of the offshore price zone is determined solely by the short-term marginal cost of offshore wind generation, which is zero, leading to a fixed CG value of zero as well.

\section{Case Study}  \label{sec:Case Study}

\begin{figure}[h]
    \centering
    \includegraphics[width=\columnwidth]{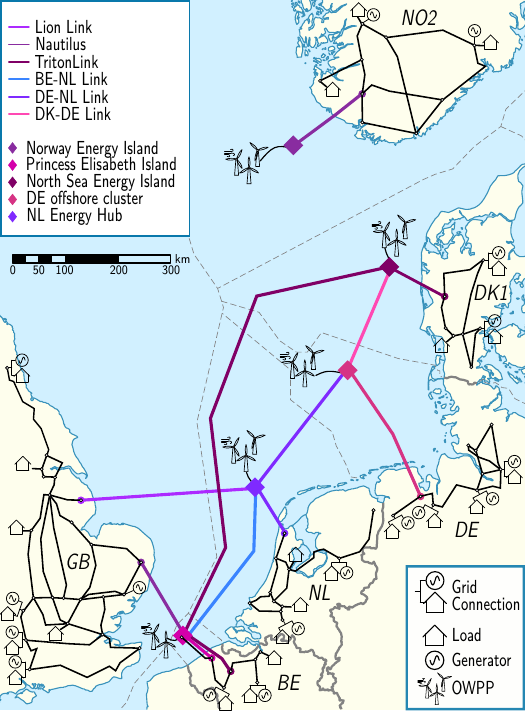}
    \caption{North Sea MTDC network}
    \label{fig:North_Sea_MTDC}
\end{figure}

This section presents a case study focused on the North Sea development area. The study explores the question: How would power flows have behaved if such developments had been in place in 2024?  The case study constructs the North Sea HVDC development area, encompassing the market areas of Belgium (BE), Denmark (DK), Germany (DE), Great Britain (GB), the Netherlands (NL), and Norway (NO). The full system is described as an AC/DC hybrid grid, where loads within the price zones are interconnected by an HVAC system. The price zones are then interconnected by an MTDC system, as shown in \autoref{fig:North_Sea_MTDC}. Supply and demand curves from the EpexSpot database \cite{epexspot} are used to calculate the $P_N$ boundaries for each price zone. Energy storage are not included within the scope of this study.

\subsection{Description}

\begin{figure}[ht]
    \centering
    \definecolor{color1}{RGB}{31, 119, 180} 
\definecolor{color2}{RGB}{255, 127, 14} 
\definecolor{color3}{RGB}{44, 160, 44} 
\definecolor{color4}{RGB}{214, 39, 40} 
\definecolor{color5}{RGB}{148, 103, 189} 
\definecolor{color6}{RGB}{140, 86, 75} 
\definecolor{color7}{RGB}{0, 191, 255} 

\footnotesize\centering
\begin{tikzpicture}
    \begin{axis}[
        width=8.5cm,
        height=3.5cm,
        grid=major,
        outer axis line style={line width=0.8pt},
        legend style={at={(0.5,1.2)}, anchor=south, legend columns=3, /tikz/every even column/.append style={column sep=0.1cm},reverse legend},
        legend image post style={line width=1.5pt, mark=none, /tikz/line cap=round},
        ylabel style={align=center, yshift=-10pt},
        ylabel={Power  [GW]},
        xmin=5750,
        xmax=6000,
        ymin=0,
        ymax=22,
        xtick={5750,5800,...,5950,6000},
        xlabel={Time step [h]}
        ]

        \addplot[color5, fill=color5, fill opacity=0.1] table [x index=0, y index=5, col sep=comma] {TS_Data/Load_dist_2024.csv} \closedcycle;
        \addlegendentry{NL}
        \addplot[color4, fill=color4, fill opacity=0.1] table [x index=0, y index=4, col sep=comma] {TS_Data/Load_dist_2024.csv} \closedcycle;
        \addlegendentry{GB}
        \addplot[color3, fill=color3, fill opacity=0.1] table [x index=0, y index=3, col sep=comma] {TS_Data/Load_dist_2024.csv} \closedcycle;
        \addlegendentry{DK}
        \addplot[color2, fill=color2, fill opacity=0.1] table [x index=0, y index=2, col sep=comma] {TS_Data/Load_dist_2024.csv} \closedcycle;
        \addlegendentry{DE}
        \addplot[color1, fill=color1, fill opacity=0.1] table [x index=0, y index=1, col sep=comma] {TS_Data/Load_dist_2024.csv} \closedcycle;
        \addlegendentry{BE}

        \addplot[blue,line width=1pt] table [x index=0, y index=7, col sep=comma] {TS_Data/Load_dist_2024.csv};
        \addlegendentry{Wind availabilty}

    \end{axis}     
\end{tikzpicture}  
    \caption{Aggregated power load by zone connected to the MTDC compared to the wind availability}
    \label{fig: TS_Load}
\end{figure}
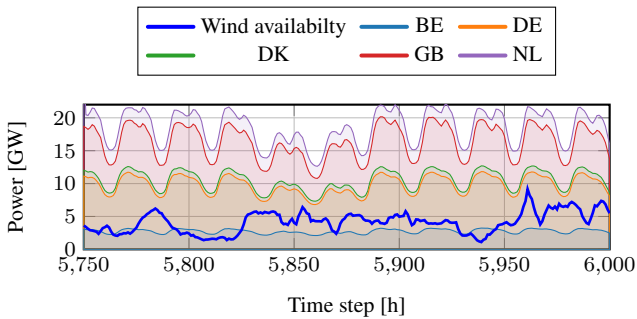

The case study aims to examine the behaviour of power generation and curtailment in offshore developments under price zone constraints. The offshore developments considered in the study are: Princess Elizabeth Island (PEI) in BE with an installed power of 3.5 GW, NL Energy Hub (NLH) with 2 GW, German Offshore Interconnection Cluster (GOC) with 4 GW, Danish Energy Island (DKEI) with 3.5 GW and a Norwegian Energy Hub (NOH) with 2 GW. A detailed description of the installed power, generation capacity, and load for each offshore wind price zone is provided in \autoref{tab:POwerSummary}. A visual representation of the load distribution across the price zones is shown in \autoref{fig: TS_Load}, alongside a comparison with the total wind power availability. The study is structured into three scenarios to facilitate comparison. Three indicators are used to evaluate the scenarios: the price in each price zone, the curtailment of offshore wind developments, and the power exchanges between price zones.

\begin{table}[htbp]
\centering
\caption{Power capacity in each market \cite{a2024_expert,a2024_entsoedata}}
\label{tab:POwerSummary}

\resizebox{0.8\columnwidth}{!}{%
\begin{tabular}{cccc}
\hline
\begin{tabular}[c]{@{}c@{}}Price\\ Zone\end{tabular} &
  \begin{tabular}[c]{@{}c@{}}Wind capacity\\ {[}MW{]}\end{tabular} &
  \begin{tabular}[c]{@{}c@{}}Other generation capacity\\ {[}MW{]}\end{tabular} &
  \begin{tabular}[c]{@{}c@{}}Load\\ {[}MW{]}\end{tabular} \\ \hline
BE & 3500  & 9340  & 3900 \\
DE & 4500 & 18680 & 10000 \\
DK & 3500  & 4074  & 1160 \\
GB & 0  & 19225  & 10900 \\
NL & 2000  & 9340  & 3600 \\
NO & 2000  & 4346  & 1500  \\ \hline
\end{tabular}%
  }
\end{table}

\begin{itemize}
    \item Scenario I (SI): The OPF is performed with a fixed price for power generation in each generator, determined by the price zone.
    \item Scenario II (SII): The $P_N$ boundaries in \eqref{eq: PNconstraint} are added to the OPF formulation, where the cost of generation depends on the power produced in the price zone. The price is thus determined by \eqref{eq:price_var} in the optimisation.
    \item Scenario III (SIII): Wind power availability is removed from SII.
\end{itemize}

All simulations were performed with \cite{Bernardo_Pyflow} on a Dell Vostro 15 3530, equipped with a 13th Gen Intel(R) Core(TM) i7-1355U processor (1.70 GHz) and 40 GB of RAM. The case studies covered all hourly time steps for the year 2024, with an average solving time of 0.44, 0.56 and 0.48 seconds for SI, SII and SIII respectively per time step. It is not prudent to show all the results in this paper, it has been chosen to show 250 time steps to show the applicability of the formulation. 

\subsection{Results}

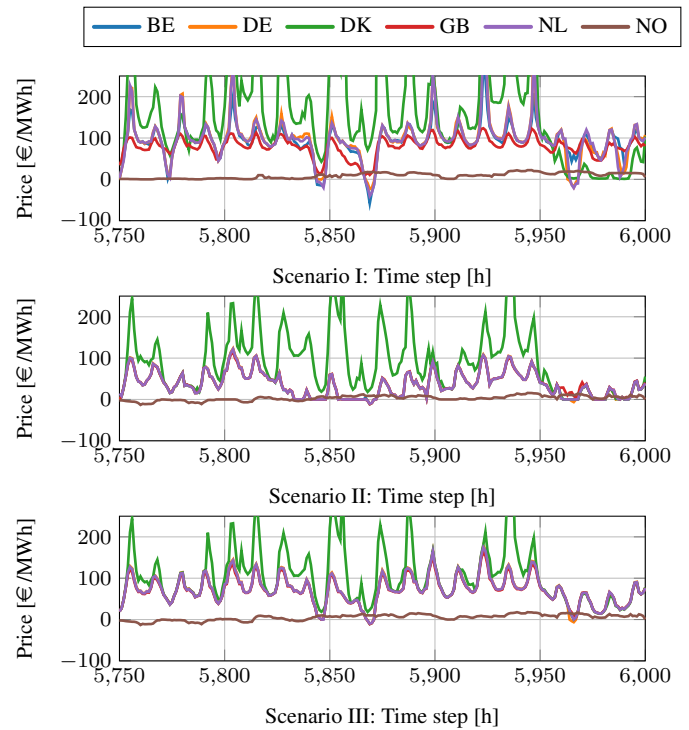
\begin{figure}[htbp]
    \centering
    \definecolor{color1}{RGB}{31, 119, 180} 
\definecolor{color2}{RGB}{255, 127, 14} 
\definecolor{color3}{RGB}{44, 160, 44} 
\definecolor{color4}{RGB}{214, 39, 40} 
\definecolor{color5}{RGB}{148, 103, 189} 
\definecolor{color6}{RGB}{140, 86, 75} 
\definecolor{color7}{RGB}{0, 191, 255} 

\footnotesize\centering
\begin{tikzpicture}
    \begin{groupplot}[
        group style={group size=1 by 3},
        width=8.5cm,
        height=3.5cm,
        grid=major,
        legend style={at={(0.5,1.2)}, anchor=south, legend columns=6, /tikz/every even column/.append style={column sep=0.1cm}},
        legend image post style={line width=1.5pt, mark=none, /tikz/line cap=round}
    ]
        \nextgroupplot[
            grid=major,
            ylabel={Price  [\EUR{}/MWh]},
            xmin=5750,
            xmax=6000,
            ymin=-100,
            ymax=250,
            xtick={5750,5800,...,5950,6000},
            xlabel={Scenario I: Time step [h]}
        ]
        \addplot[color1,line width=1pt] table [x index=0, y index=1, col sep=comma] {TS_Data/S1_Price_Data.csv} ;
        \addlegendentry{BE}
        \addplot[color2,line width=1pt] table [x index=0, y index=2, col sep=comma] {TS_Data/S1_Price_Data.csv} ;
        \addlegendentry{DE}
        \addplot[color3,line width=1pt] table [x index=0, y index=3, col sep=comma] {TS_Data/S1_Price_Data.csv} ;
        \addlegendentry{DK}
        \addplot[color4,line width=1pt] table [x index=0, y index=4, col sep=comma] {TS_Data/S1_Price_Data.csv} ;
        \addlegendentry{GB}
        \addplot[color5,line width=1pt] table [x index=0, y index=5, col sep=comma] {TS_Data/S1_Price_Data.csv} ;
        \addlegendentry{NL}
        \addplot[color6,line width=1pt] table [x index=0, y index=6, col sep=comma] {TS_Data/S1_Price_Data.csv} ;
        \addlegendentry{NO}

        \nextgroupplot[
            grid=major,
            ylabel={Price  [\EUR{}/MWh]},
            xmin=5750,
            xmax=6000,
            ymin=-100,
            ymax=250,
            xtick={5750,5800,...,5950,6000},
            xlabel={Scenario II: Time step [h]}
        ]

        \addplot[color1,line width=1pt] table [x index=0, y index=1, col sep=comma] {TS_Data/S2_Price_Data.csv} ;
        \addplot[color2,line width=1pt] table [x index=0, y index=2, col sep=comma] {TS_Data/S2_Price_Data.csv} ;
        \addplot[color3,line width=1pt] table [x index=0, y index=3, col sep=comma] {TS_Data/S2_Price_Data.csv} ;
        \addplot[color4,line width=1pt] table [x index=0, y index=4, col sep=comma] {TS_Data/S2_Price_Data.csv} ;
        \addplot[color5,line width=1pt] table [x index=0, y index=5, col sep=comma] {TS_Data/S2_Price_Data.csv} ;
        \addplot[color6,line width=1pt] table [x index=0, y index=6, col sep=comma] {TS_Data/S2_Price_Data.csv} ;
       
        \nextgroupplot[
            grid=major,
            ylabel={Price  [\EUR{}/MWh]},
            xmin=5750,
            xmax=6000,
            ymin=-100,
            ymax=250,
            xtick={5750,5800,...,5950,6000},
            xlabel={Scenario III: Time step [h]}
        ]

        \addplot[color1,line width=1pt] table [x index=0, y index=1, col sep=comma] {TS_Data/S3_Price_Data.csv} ;
        \addplot[color2,line width=1pt] table [x index=0, y index=2, col sep=comma] {TS_Data/S3_Price_Data.csv} ;
        \addplot[color3,line width=1pt] table [x index=0, y index=3, col sep=comma] {TS_Data/S3_Price_Data.csv} ;
        \addplot[color4,line width=1pt] table [x index=0, y index=4, col sep=comma] {TS_Data/S3_Price_Data.csv} ;
        \addplot[color5,line width=1pt] table [x index=0, y index=5, col sep=comma] {TS_Data/S3_Price_Data.csv} ;
        \addplot[color6,line width=1pt] table [x index=0, y index=6, col sep=comma] {TS_Data/S3_Price_Data.csv} ;

    \end{groupplot}     
\end{tikzpicture}  
    \caption{Electricity price in different markets \cite{EPEXSpotBasics}}
    \label{fig: Price}
\end{figure}

In SI, the price is treated as a parameter in the formulation, whereas in SII, it is treated as a variable. The resulting prices for both scenarios are shown in \autoref{fig: Price}. In SI, the price shown is the input for the optimisation and remains unaffected by power exchanges. In contrast, in SII, prices are influenced by power exchanges and adjust dynamically based on the optimisation results. The increased availability of zero-priced energy, from offshore wind developments, causes prices in all price zones to decrease. If prices are negative, such as $t\in[5867,5871]$, power export is prioritised.This is because the cost of generation becomes minimal at zero price, thereby reducing the magnitude of the negative prices. In SIII, when there is no wind power input, the MTDC operates solely in its interconnector functionality. In \autoref{fig: Price}, the prices of interconnected price zones equilibrate for each time period. Since NO is not connected to the MTDC, its prices remain relatively unaffected.

\begin{figure}[htbp]
    \centering
    \definecolor{color1}{RGB}{31, 119, 180} 
\definecolor{color2}{RGB}{255, 127, 14} 
\definecolor{color3}{RGB}{44, 160, 44} 
\definecolor{color4}{RGB}{214, 39, 40} 
\definecolor{color5}{RGB}{148, 103, 189} 
\definecolor{color6}{RGB}{0, 128, 128} 
\definecolor{color7}{RGB}{227, 119, 194} 
\footnotesize\centering
\begin{tikzpicture}
    \begin{groupplot}[
        group style={group size=1 by 3},
        width=8.5cm,
        height=3.5cm,
        grid=major,
        legend style={at={(0.5,1.2)}, anchor=south, legend columns=5, /tikz/every even column/.append style={column sep=0.1cm}},
        legend image post style={line width=1.5pt, mark=none, /tikz/line cap=round},
        ylabel style={align=center, yshift=-10pt}
    ]
        \nextgroupplot[
            grid=major,
            ylabel={Curtialment [\%]},
            xmin=5750,
            xmax=6000,
            xtick={5750,5800,...,5950,6000},
            xlabel={Scenario I: Time step [h]}
        ]
        \addplot[color1,line width=1pt] table [x index=0, y index=1, col sep=comma] {TS_Data/S1_curtailment_data.csv};
        \addlegendentry{PEI}
        \addplot[color2,line width=1pt] table [x index=0, y index=2, col sep=comma] {TS_Data/S1_curtailment_data.csv};
        \addlegendentry{GOC}       
        \addplot[color3,line width=1pt] table [x index=0, y index=3, col sep=comma] {TS_Data/S1_curtailment_data.csv};
        \addlegendentry{DKEI}
        \addplot[color5,line width=1pt] table [x index=0, y index=4, col sep=comma] {TS_Data/S1_curtailment_data.csv};
        \addlegendentry{NLH}
        \addplot[color6,line width=1pt] table [x index=0, y index=5, col sep=comma] {TS_Data/S1_curtailment_data.csv};
        \addlegendentry{NOH}

     \nextgroupplot[
            grid=major,
            ylabel={Curtialment [\%]},
            xmin=5750,
            xmax=6000,
            xtick={5750,5800,...,5950,6000},
            xlabel={Scenario II: Time step [h]}
        ]
         \addplot[color1,line width=1pt] table [x index=0, y index=1, col sep=comma] {TS_Data/S2_curtailment_data.csv};
       
        \addplot[color2,line width=1pt] table [x index=0, y index=2, col sep=comma] {TS_Data/S2_curtailment_data.csv};
        
        \addplot[color3,line width=1pt] table [x index=0, y index=3, col sep=comma] {TS_Data/S2_curtailment_data.csv};
      
        \addplot[color5,line width=1pt] table [x index=0, y index=4, col sep=comma] {TS_Data/S2_curtailment_data.csv};
        \addplot[color6,line width=1pt] table [x index=0, y index=5, col sep=comma] {TS_Data/S2_curtailment_data.csv};

    \end{groupplot}     
\end{tikzpicture}  
    \caption{Offshore wind curtailment in different markets}
    \label{fig: TS_curt}
\end{figure}
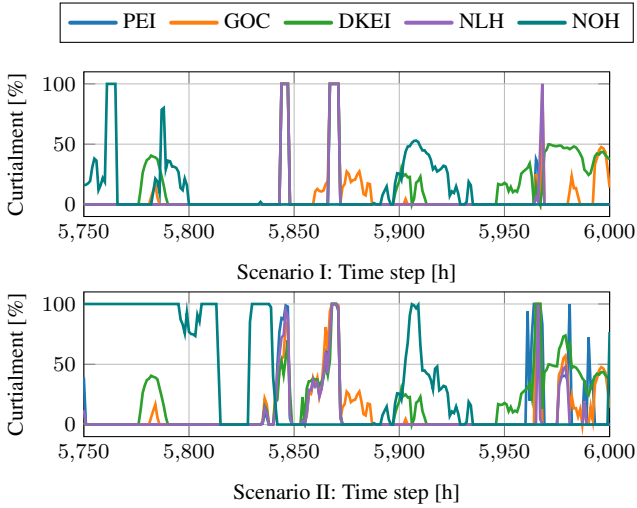

Limiting the power that each price zone can import significantly impacts the curtailment of offshore wind developments. \autoref{fig: TS_curt} illustrates the progression of curtailment in both scenarios. In SI, where there is no limit on imports in price zones, curtailment is minimal. Curtailment only occurs if a price zone experiences negative prices or if offshore wind generation exceeds the local load  (e.g. $t \in [5898,5928] \in m = NO$). Looking back at \autoref{fig: TS_Load}, we observe that curtailment due to excess offshore wind generation is highly unlikely for the remaining price zones connected by the MTDC. Therefore, the observed curtailment in SI is attributed to price zones prioritising negative prices over wind resources or is limited by the power flows in the MTDC (e.g. $t \in [5968,6000] \in rg = DKEI$).

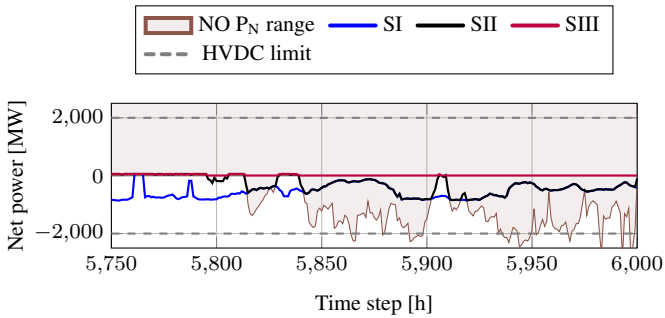
\begin{figure}[htbp]
    \centering
    \definecolor{color1}{RGB}{31, 119, 180} 
\definecolor{color2}{RGB}{255, 127, 14} 
\definecolor{color3}{RGB}{44, 160, 44} 
\definecolor{color4}{RGB}{214, 39, 40} 
\definecolor{color5}{RGB}{148, 103, 189} 
\definecolor{color6}{RGB}{140, 86, 75} 
\definecolor{color7}{RGB}{0, 191, 255} 

\footnotesize\centering
\begin{tikzpicture}
    \begin{axis}[
        width=8.5cm,
        height=3.5cm,
        grid=major,
        legend style={at={(0.5,1.2)}, anchor=south, legend columns=4, /tikz/every even column/.append style={column sep=0.1cm}},
        legend image post style={line width=1.5pt, mark=none, /tikz/line cap=round},
        ylabel={Net power  [MW]},
        xmin=5750,
        xmax=6000,
        ymin=-2500,
        ymax=2500,
        xtick={5750,5800,...,5950,6000},
        xlabel={Time step [h]}
    ]
        \addplot[name path=upper, color6, forget plot] table [x index=0, y index=26, col sep=comma] {TS_Data/PN_2024.csv};
        \addplot[name path=lower, color6, line width=0.5pt, forget plot] table [x index=0, y index=27, col sep=comma] {TS_Data/PN_2024.csv};
        \addplot[color6, fill=color6, fill opacity=0.1,line width =0.5pt] fill between[of=upper and lower];
        \addlegendentry{NO P\textsubscript{N} range}

        \addplot[blue, line width=0.8pt] table [x index=0, y index=28, col sep=comma] {TS_Data/PN_2024.csv};
        \addlegendentry{SI}
        
        \addplot[black, line width=0.8pt] table [x index=0, y index=29, col sep=comma] {TS_Data/PN_2024.csv};
        \addlegendentry{SII}

        \addplot[purple, line width=0.8pt] table [x index=0, y index=30, col sep=comma] {TS_Data/PN_2024.csv};
        \addlegendentry{SIII}
        
        \addplot[dashed, thick, color=gray] coordinates {(5750, 2000) (6000, 2000)};
        \addplot[dashed, thick, color=gray] coordinates {(5750, -2000) (6000, -2000)};
        \addlegendentry{HVDC limit}
        
    \end{axis}     
\end{tikzpicture}  
    \caption{Power exchange in price zone NO}
    \label{fig: TS_PN_NO}
\end{figure}

\begin{figure}[htbp]
    \centering
    \definecolor{color1}{RGB}{31, 119, 180} 
\definecolor{color2}{RGB}{255, 127, 14} 
\definecolor{color3}{RGB}{44, 160, 44} 
\definecolor{color4}{RGB}{214, 39, 40} 
\definecolor{color5}{RGB}{148, 103, 189} 
\definecolor{color6}{RGB}{140, 86, 75} 
\definecolor{color7}{RGB}{0, 191, 255} 

\footnotesize\centering
\begin{tikzpicture}
    \begin{axis}[
        width=8.5cm,
        height=3.5cm,
        grid=major,
        outer axis line style={line width=0.8pt},
        legend style={at={(0.5,1.2)}, anchor=south, legend columns=3, /tikz/every even column/.append style={column sep=0.1cm},reverse legend},
        legend image post style={line width=1.5pt, mark=none, /tikz/line cap=round},
        ylabel style={align=center, yshift=-10pt},
        ylabel={Power  [GW]},
        xmin=5750,
        xmax=6000,
        ymin=0,
        xtick={5750,5800,...,5950,6000},
        xlabel={Time step [h]}
        ]
   
        \addplot[color5, fill=color5, fill opacity=0.1] table [x index=0, y index=5, col sep=comma] {TS_Data/imoport_max_2024.csv} \closedcycle;
        \addlegendentry{NL}
        \addplot[color4, fill=color4, fill opacity=0.1] table [x index=0, y index=4, col sep=comma] {TS_Data/imoport_max_2024.csv} \closedcycle;
        \addlegendentry{GB}
        \addplot[color3, fill=color3, fill opacity=0.1] table [x index=0, y index=3, col sep=comma] {TS_Data/imoport_max_2024.csv} \closedcycle;
        \addlegendentry{DK}
        \addplot[color2, fill=color2, fill opacity=0.1] table [x index=0, y index=2, col sep=comma] {TS_Data/imoport_max_2024.csv} \closedcycle;
        \addlegendentry{DE}
        \addplot[color1, fill=color1, fill opacity=0.1] table [x index=0, y index=1, col sep=comma] {TS_Data/imoport_max_2024.csv} \closedcycle;
        \addlegendentry{BE}

        \addplot[blue,line width=1pt] table [x index=0, y index=7, col sep=comma] {TS_Data/imoport_max_2024.csv};
        \addlegendentry{Wind availabilty}

    \end{axis}     
\end{tikzpicture}  
    \caption{Aggregated power import limit by zone connected to the MTDC compared to the wind availability} 
    \label{fig: TS_import}
\end{figure}
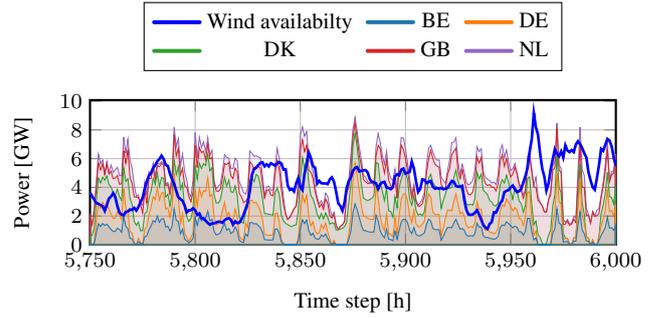

In contrast, SII introduces curtailment not only under the conditions of negative pricing but also when the import capacities of price zones are fully utilised such as $t \in [5750,5797] \in m=NO$ as shown in \autoref{fig: TS_PN_NO} . To better understand where the additional curtailment arises in SII, \autoref{fig: TS_import} provides a visual guide to the situation. This figure highlights that the primary limiting factor is the import constraint. Curtailment increases when the available wind energy exceeds the aggregate import capabilities of all connected price zones (e.g. $t \in [5973,5994]$). 

\begin{figure}[htbp]
    \centering
    \definecolor{color1}{RGB}{31, 119, 180} 
\definecolor{color2}{RGB}{255, 127, 14} 
\definecolor{color3}{RGB}{44, 160, 44} 
\definecolor{color4}{RGB}{214, 39, 40} 
\definecolor{color5}{RGB}{148, 103, 189} 
\definecolor{color6}{RGB}{140, 86, 75} 
\definecolor{color7}{RGB}{0, 191, 255} 

\footnotesize\centering
\begin{tikzpicture}
    \begin{axis}[
        width=8.5cm,
        height=3.5cm,
        grid=major,
        legend style={at={(0.5,1.2)}, anchor=south, legend columns=4, /tikz/every even column/.append style={column sep=0.1cm}},
        legend image post style={line width=1.5pt, mark=none, /tikz/line cap=round},
        ylabel={Net power  [MW]},
        xmin=5750,
        xmax=6000,
        ymin=-2500,
        ymax=2500,
        xtick={5750,5800,...,5950,6000},
        xlabel={Time step [h]}
    ]
        \addplot[name path=upper, color5, forget plot] table [x index=0, y index=21, col sep=comma] {TS_Data/PN_2024.csv};
        \addplot[name path=lower, color5, line width=0.5pt, forget plot] table [x index=0, y index=22, col sep=comma] {TS_Data/PN_2024.csv};
        \addplot[color5, fill=color5, fill opacity=0.1,line width =0.5pt] fill between[of=upper and lower];
        \addlegendentry{NL P\textsubscript{N} range}

        \addplot[blue, line width=0.8pt] table [x index=0, y index=23, col sep=comma] {TS_Data/PN_2024.csv};
        \addlegendentry{SI}
        
        \addplot[black, line width=0.8pt] table [x index=0, y index=24, col sep=comma] {TS_Data/PN_2024.csv};
        \addlegendentry{SII}

        \addplot[purple, line width=0.8pt] table [x index=0, y index=25, col sep=comma] {TS_Data/PN_2024.csv};
        \addlegendentry{SIII}
        
        \addplot[dashed, thick, color=gray] coordinates {(5750, 2000) (6000, 2000)};
        \addplot[dashed, thick, color=gray] coordinates {(5750, -2000) (6000, -2000)};
        \addlegendentry{MTDC limit}
        
    \end{axis}     
\end{tikzpicture}  
    \caption{Power exchange in price zone NL}
    \label{fig: TS_PN_NL}
\end{figure}
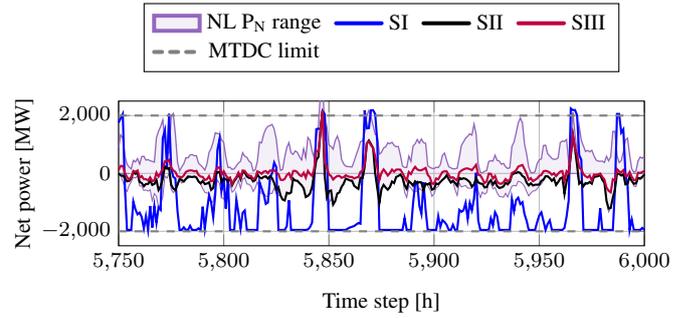

Taking a small price zone such as NL in \autoref{fig: TS_PN_NL} as an example, when we compare the value of $P_N$. In SI, there is an excess of imports in most cases. However, during time steps with negative price values (e.g. $t \in [5867,5871]$) the export power in SI is maximized, reaching the interconnector limit of 2 GW. These situations are unrealistic in day-to-day operations, as excessive exports would cause market prices to skyrocket. On the other hand, excessive imports would result in negative prices in the day-ahead market. In SII, with wind power available to import, in most cases $P_N$ results in the maximum import level. In SIII, in the aforementioned range, NL has a lower price compared to other zones and would then export power.

\begin{figure}[htbp]
    \centering
    \definecolor{color1}{RGB}{31, 119, 180} 
\definecolor{color2}{RGB}{255, 127, 14} 
\definecolor{color3}{RGB}{44, 160, 44} 
\definecolor{color4}{RGB}{214, 39, 40} 
\definecolor{color5}{RGB}{148, 103, 189} 
\definecolor{color6}{RGB}{140, 86, 75} 
\definecolor{color7}{RGB}{0, 191, 255} 

\footnotesize\centering
\begin{tikzpicture}
    \begin{axis}[
        width=8.5cm,
        height=3.5cm,
        grid=major,
        legend style={at={(0.5,1.2)}, anchor=south, legend columns=4, /tikz/every even column/.append style={column sep=0.1cm}},
        legend image post style={line width=1.5pt, mark=none, /tikz/line cap=round},
        ylabel={Net power  [MW]},
        xmin=5750,
        xmax=6000,
        ymax=4500,
        xtick={5750,5800,...,5950,6000},
        xlabel={Time step [h]}
    ]
        \addplot[name path=upper, color4, forget plot] table [x index=0, y index=16, col sep=comma] {TS_Data/PN_2024.csv};
        \addplot[name path=lower, color4, line width=0.5pt, forget plot] table [x index=0, y index=17, col sep=comma] {TS_Data/PN_2024.csv};
        \addplot[color4, fill=color4, fill opacity=0.1,line width =0.5pt] fill between[of=upper and lower];
        \addlegendentry{GB P\textsubscript{N} range}

        \addplot[blue, line width=0.8pt] table [x index=0, y index=18, col sep=comma] {TS_Data/PN_2024.csv};
        \addlegendentry{SI}
        
        \addplot[black, line width=0.8pt] table [x index=0, y index=19, col sep=comma] {TS_Data/PN_2024.csv};
        \addlegendentry{SII}

        \addplot[purple, line width=0.8pt] table [x index=0, y index=20, col sep=comma] {TS_Data/PN_2024.csv};
        \addlegendentry{SIII}
        
        \addplot[dashed, thick, color=gray] coordinates {(5750, 3400) (6000, 3400)};
        \addplot[dashed, thick, color=gray] coordinates {(5750, -3400) (6000, -3400)};
        \addlegendentry{MTDC limit}
    
    \end{axis}     
\end{tikzpicture}  
    \caption{Power exchange in price zone GB}
    \label{fig: TS_PN_GB}
\end{figure}
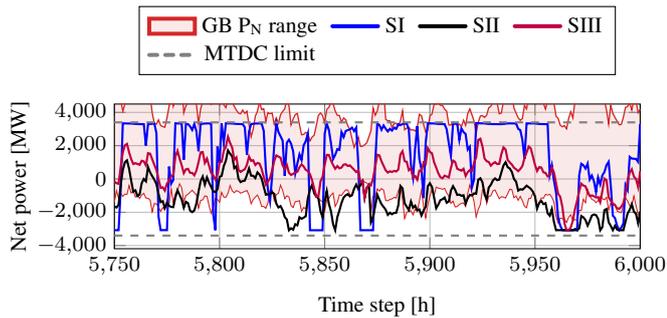

Considering a large price zone, such as GB, connected to the MTDC via the interconnectors Nautilus (1.4 GW) and Lion Link (2 GW). In \autoref{fig: TS_PN_GB}, for SI, the import value reaches its maximum of 3.4 GW, constrained by the combined capacities of the interconnectors. The $P_N$ range is comparatively larger for GB than the smaller price zone NL. In other words, imports and exports have a less significant effect on the price. Most export flows in SI would still be considered within the limits set by $P_N$ but now are limited by the MTDC rated power (e.g. $t\in [5922,5956]$), However, in SIII for the same range of time frames prices have reached an equilibrium at a reduced value of $P_N$ and exporting more power is not viable. The purpose of this figure is to show that constraining $P_N$ alone is insufficient; more importantly,  a relationship between $P_N$ and the price or cost of generation in a price zone is required.

\section{Conclusions}

This paper expands the formulation of non-linear AC/DC optimal power flow with inclusion of transmission constraints between price zones. The formulation allows for steady state analysis of imports and exports of the integrated price zones and how these impact the equilibrium prices. Each price zone could be modelled as singular nodes or, more importantly, as collections of AC and DC nodes, as presented in the formulation.

The model maximizes the social welfare by assuming constant benefit of consumption and minimizing the cost of generation by optimising the curtailment of renewable energy sources and the power set points of converters and generators. The optimisation considers loading inputs, renewable energy source availability, and a quadratic relationship between the exchanged power in a price zone and its cost of generation for a given instance. 

The 2030 development plan for the North Sea Energy Hubs is used as a case study to exemplify the proposed formulation. Three cases are compared: the first assumes the market equilibrium price as a fixed parameter in the optimisation, while the second treats the price as a variable dependent on market exchanges, and the third eliminates wind power availability to more easily visualise power exchanges between price zones. The case study demonstrates that the first scenario fails to accurately capture power exchanges, as large power flows become price-makers in the market. Therefore, additional constraints beyond line capacity limits must be imposed to address this limitation.

This extension to the optimisation formulation is expected to support further studies on hybrid AC/DC grids, including, but not limited to, transmission expansion planning.

\section{Acknowledgements}
This work has received funding from the ADOreD project of the European Union’s Horizon Europe Research and Innovation program under the Marie Skłodowska-Curie grant
agreement No 101073554.
\FloatBarrier
\section{References}

\bibliographystyle{IEEEtran}
\bibliography{sources}

\end{document}